\documentclass[12pt]{iopart}

\usepackage{graphicx}
\begin{document}

\title{Risk, ambiguity and quantum decision theory}

\author{Riccardo Franco
\footnote[3]{To whom correspondence should be addressed riccardo.franco@polito.it}}

\date{\today}

\begin{abstract}
In the present article we use the quantum formalism to describe the effects of risk and ambiguity in decision theory. The main idea is that the probabilities in the classic theory of expected utility are estimated probabilities, and thus do not follow the classic laws of probability theory. In particular, we show that it is possible to use consistently the classic expected utility formula, where the probability associated to the events are computed with the equation of quantum interference. Thus we show that the correct utility of a lottery can be simply computed by adding to the classic expected utility a new corrective term, the uncertainty utility, directly connected with the quantum interference term. 
\end{abstract}

\maketitle

%
\section{Introduction}
This article is a first attempt to provide a consistent description of the main effects of decision theory within the quantum formalism. A number of attempts have been done to apply the formalism of quantum mechanics to domains of science different from the micro-world with applications to economics, operations research and management science, psychology and cognition, game theory, and language and artificial intelligence. For a list of references, see \cite{Rfranco_rat_ign_3, Rfranco_rconj}. However, these attempts are very different and they do not provide a general way to apply the quantum formalism. 
In \cite{Mura1} there is a quantum-like approach to describe risk and uncertainty, thought its interpretation is quite difficult. 

Our approach is based on the fact that classic expected theory can be modified by imposing to the judged probabilities different laws. This leads to a formula very similar to the classic expected utility equation, with an additional term, whose interpretation is clear and simple.
In recent works, it has been shown that the judged probabilities
seem to be adequately described by a quantum formalism. In
particular, the inverse fallacy \cite{Rfranco_rat_ign_3} has been shown to be a direct consequence of the quantum formalism. Moreover, the conjunction
fallacy \cite{Rfranco_rconj} can be described in terms of the quantum interference, leading to a good agreement with experimental results.

The judged probability may not be the only object compatible with the quantum
framework: in the present article, we want to show that also the preferences, and thus
the expected utilities, could be described  in quantum terms in a very simple way, by following the same definitions of the expected utility theory  \cite{Neumann}. In particular, we will show that the quantum formalism is able to give a consistent description of the
effects of risk and ambiguity, evidenced for example in the two-color Ellsberg's
experiment \cite{ellsberg}.
Before Ellsberg's work, Keynes \cite{Keynes} observed that people's willingness to act
in the presence of uncertainty depends on the perceived probability of the event in
question. The Ellsberg's work tried to generalize this fact, by defining the concept
of vagueness or ambiguity as \textit{a quality depending on the amount, type, and
'unanimity' of information, and giving rise to one's degree of 'confidence' in an
estimate of relative likelihoods}.
The ambiguity effects \cite{Tv_Fox1, Fox_Weber} have been demonstrated in numerous settings, such as laboratory choice
experiments \cite{Curley_Yates}, market experiments \cite{Sarin_Weber}, and in
contextualized decisions \cite{Ho_Keller}. Ambiguity aversion
 is referred to as one of the most prominent violations of the expected utility theory.
%
%
\section{The expected utility theory}
The von Neumann-Morgensten expected utility model \cite{Neumann} provides the
conceptual and computational framework that is most often used to
analyze decisions: uncertainty
about the future is represented by the following primitive
objects:
\begin{enumerate}
\item the decision maker (DM), also called the agent
\item
a set $S$ of \textit{consequences} or \textit{possible  outcomes} for the decision
maker. They could be amounts of money in the bank or more general states of the person
such as health, happiness, pleasant or unpleasant experiences, and so on. We can write
explicitly a finite set of mutually exclusive and exhaustive outcomes $(s_1,...,
s_n)$. 
\item
a set $\Delta$ of \textit{ probability distributions} or \textit{lotteries}, defined
over the possible outcomes. A lottery is the assignment $(P_1,..., P_n)$, where $P_i$
is the probability relevant to the outcome $s_i$. 
\item
a \textit{preference relation} $\geq$ over the lotteries, which
characterizes the decision maker. The statement $f \geq f'$ has
the following interpretation: the lottery $f$ is at least
as desirable as  $f'$. Given $\geq$, the
strict preference relation $>$ is defined as $f
> f'$ iff $f \geq f'$ and not $f' \geq f$, while the indifference
relation $\sim$  as  $f \sim f'$ iff $f \geq f'$ and $f' \geq f$.
\end{enumerate}
The preference ordering is assumed to satisfy the following axioms:
\begin{enumerate}
\item Completeness: $\forall f,g\in \Delta$: or $f\geq g$ or $g\geq f$
\item Transitivity: $\forall f,g,h\in \Delta$: if $f\geq g$ and $g\geq h$, then $f\geq h$
\item Continuity: $\forall f,g,h\in \Delta$ with $f\geq g \geq h$, there exists $\alpha,\beta\in(0,1)$ such that $\alpha f +(1-\alpha)h\geq g\geq \beta f+ (1-\beta)h$
\item Independence: $\forall f,g,h\in \Delta$ we have that $f\geq g$ if and only if $\forall\alpha\in(0,1)$ we have that $\alpha f +(1-\alpha)h\geq \alpha g+(1-\alpha)h$
\end{enumerate}
The axioms previously defined allow to show that relation $f \geq f'$ is equivalent to
the inequality $u(f)\geq u(f')$, where $u(f)$ is a functional $\Delta \rightarrow R$
called the utility function:
\begin{equation}\label{EU}
u(f)= \sum_{i=1}^{n}P_i u(s_i).
\end{equation}
Thus the utility associated to a lottery $\{P_i\}$ can be computed after assigning to
any outcome $s_i$ a real value $u(s_i)$.  This recipe for rational decision making has
ancient roots: it was first proposed by Daniel Bernoulli (1738) to explain aversion to
risk in problems of gambling and insurance as well as to solve the famous St.
Petersburg Paradox. The idea of seeking to maximize the expected value of a utility
function was discarded by later generations of economists, who doubted that utility
could ever be measured on a cardinal numerical scale. However, it was later revived and
rehabilitated, evidencing that the expected-utility model could be derived from simple
and seemingly reasonable axioms of consistent preferences under risk and uncertainty,
in which a pivotal role is played by an independence condition known as the sure-thing
principle.

Von Neumann and Morgenstern  \cite{Neumann} consider the special case in which states of the world
have  objectively known probabilities (as in games of chance), while Savage extends
the model to include situations where probabilities are subjectively determined by the
decision maker.
%
%
\section{Judged probabilities}
We introduce some very simple definitions, which will be used both in the classic and in the quantum framework to describe the estimated probabilities and the preferences.
\begin{itemize}
\item An \textit{observable} $A$ is an event which can be verified. For any observable $A$ we can
always write the dichotomous question "is $A$ true?". In the following, we identify
the event $a_1$ with the answer Yes, and we call $a_0$ the answer No, or the negation
of the event.
\item The \textit{preparation} is any information previously given to the agent
which can be used to determine the estimated probabilities.
\item The \textit{opinion state} (or simply state) of an agent is the result of the preparation.
\item $P(a_1)$ is the \textit{estimated probability} that the event is true, given a set of agents
in the same opinion state. Analogously, we call
$P(a_0)$ the estimated probability that the event is false. Of course, $P(a_0)+P(a_1)=1$.
\end{itemize}

\subsection{Classic judged probabilities}
The main hypothesis of this section is that the judged (or estimated) probability follows the same rules of the classic
theory of probability.
For any observable $A$, and thus for any couple of mutually exclusive events $(a_0, a_1)$, the opinion state of agents can be described as the vector
\begin{equation}\label{vectorA}
\left[\begin{tabular}{c}
$P(a_0)$\\
$P(a_1)$\\
\end{tabular}\right]\,,
\end{equation}
which describes the judged probabilities associated to the events: such probabilities can be estimated by a single agent, or they can be the mean value of probabilites estimated by a set of agents in the same preparation.
Let us now consider a different couple of mutually exclusive events $(b_0, b_1)$,
for which we have the vector of judged probabilities
\begin{equation}\label{vectorB}
\left[\begin{tabular}{c}
$P(b_0)$\\
$P(b_1)$\\
\end{tabular}\right]\,.
\end{equation}
In the simplest case, vector (\ref{vectorB}) can be written in terms of vector (\ref{vectorA}) by means of
the \textit{transition probability matrix}, defined as
the square matrix whose elements are
\begin{equation}\label{bayes}
T(i,j)=P(b_j|a_i) \,,\,\, i,j=0,1\,\,,
\end{equation}
where $P(b_j|a_i)$ are the judged conditional probabilities.
We can express the Bayes' rule in terms of this matrix
$P(b_j)=\sum_{i}T(i,j)P(a_i)$, or equivalently
\begin{equation}\label{bayes}
\left[\begin{tabular}{c}
$P(b_0)$\\
$P(b_1)$\\
\end{tabular}\right]
=
\left[\begin{tabular}{c c}
$P(b_0|a_0)$ & $P(b_0|a_1)$\\
$P(b_1|a_0)$ & $P(b_1|a_1)$\\
\end{tabular}\right]
\left[\begin{tabular}{c}
$P(a_0)$\\
$P(a_1)$\\
\end{tabular}\right]\,.
\end{equation}
From this equation, and from the normalization of probability distribution $P(b_j)$,
we have in general that $\sum_{j}T(i,j)=1$: this defines a \textit{right stochastic matrix}, which is a
square matrix each of whose rows consists of nonnegative real numbers, with each row summing to 1.
On the contrary, a \textit{doubly stochastic matrix} is a square matrix for which all entries are
nonnegative and all rows and all columns sum to 1.
We recall the important fact that a right stochastic matrix for which $T(i,j)=T(j,i)$
is also a doubly stochastic matrix.
\subsection{Quantum judged probabilities}
First, we recall that in the quantum
formalism the probability state about the decision of a player is not fully described by probability distributions
$P(a_i)$ or $P(b_j)$. We introduce the quantum formalism, through the following rules:
\begin{itemize}
\item The opinion state $s$ about the observable $A$ in a particular preparation is completely described by the following vector, called the \textit{amplitude vector}.
\begin{equation}\label{opinion_state}
\left [\begin{tabular}{c}
$\sqrt{P(a_0)} e^{i\phi_0}$\\
$\sqrt{P(a_1)} e^{i\phi_1}$
\end{tabular}\right]
\end{equation}
Of course, the estimated probability $P(a_i)$ relevant to the event $a_i$ is the
square modulus of the corresponding element  in the couple describing the opinion
state. Thus the quantum approach differs from the classic in that there are two new
parameters, the phases $e^{i\phi_j}$, with $j=0,1$.
\item
Given a second observable $B$ with corresponding events $b_0$ and $b_1$, the opinion
state about $B$ can be written as the vector
\begin{equation}\label{opinion_state_B}
\left [\begin{tabular}{c}
$\sqrt{P(b_0)} e^{i\psi_0}$\\
$\sqrt{P(b_1)} e^{i\psi_1}$
\end{tabular}\right]
\end{equation}
Moreover, the judged conditional probabilities $P(b_j|a_i)$ are such that
\begin{equation}\label{conditional2}
P(b_j|a_i)=P(a_i|b_j)
\end{equation}
for all $i,j$. This fact is always true in the quantum formalism, while it is not
valid in general in classic probability theory \cite{Rfranco_rat_ign_3}. If the
estimated conditional probabilities follow equation (\ref{conditional2}), we have the
inverse fallacy.
\item Given a preparation of the opinion state, we can describe it in terms of question $A$ or $B$.
The equation which links the estimated probability $P(b_j)$ to $P(a_i)$ can not be
written with formula (\ref{bayes}). The quantum version of this equation is
\begin{equation}\label{transition2}
e^{i\psi_j}\sqrt{P(b_j)}=\sum_{i}U_{i,j}\sqrt{P(a_i)}e^{i\phi_i}\,,
\end{equation}
where $U_{i,j}$ is a unitary matrix. In the following, we will not use all the parameters describing a unitary matrix, and thus we will use this simplified unitary matrix
\begin{equation}\label{transition2}
  U = \left [
  \begin{tabular}{cc}
    $\sqrt{P(b_0|a_0)}$ & $-\sqrt{P(b_0|a_1)}$ \\
    $\sqrt{P(b_1|a_0)}$ & $\sqrt{P(b_1|a_1)}$
    \end{tabular} \right ] \,
\end{equation}
Matrix $U$ has the same function of the classic transition probability matrix, but it applies to the quantum opinion state (\ref{opinion_state}).
The opinion state about $B$ thus can be written in terms of the vector of amplitudes
\begin{equation}\label{opinion_state_B}
\left[\begin{tabular}{c}
$\sqrt{P(a_0)P(b_0|a_0)}e^{i\phi_0}-
\sqrt{P(a_1)P(b_0|a_1)}e^{i\phi_1}$\\
$\sqrt{P(a_0)P(b_1|a_0)}e^{i\phi_0}+
\sqrt{P(a_1)P(b_1|a_1)}e^{i\phi_1}$
\end{tabular}
\right]
\end{equation}
Thus the analogue of Bayes' rule for $P(b_1)$ in the quantum formalism is obtained with the square modulus of the 
elements of vector (\ref{opinion_state_B}):
\begin{equation}\label{interference}
\begin{tabular}{c}
$P(b_1)=P(a_0)P(b_1|a_0)+P(a_1)P(b_1|a_1)+$\\
$+2\sqrt{P(a_0)P(a_1)P(b_1|a_0)P(b_1|a_1)}cos(\phi_0-\phi_1)$\,,
\end{tabular}
\end{equation}
where the additional term can be named the \textit{interference term} $I$, and its
sign depends on the sign of $cos(\phi_0-\phi_1)$. In \cite{Rfranco_rconj} it is
shown that strongly negative interference terms can explain the conjunction fallacy.
\end{itemize}
In the most general case, a quantum state is described by  a \textit{mixed state},
that is a state for which the preparation is not completely determined. For example,
the state may be in the preparation $1$ with a probability $P_1$, and in the
preparation $2$ with a probability $P_2$ (the two vectors may  be not orthogonal). The
formal mathematical description of mixed  states leads to the formalism of density
matrix \cite{Rfranco_rconj}. Here we only note that two different complete
preparations may differ in the following three parameters: $P(a_0),P(a_1),
\phi_1-\phi_0$ (hence, in the interference term $I$). Thus the generalization of
equation (\ref{interference}) for mixed states is
\begin{equation}\label{interference_mixed}
P(b_1)=\sum_k P_k [P_k(a_0)P(b_1|a_0)+ P_k(a_1)P(b_1|a_1)+I_k ]
\end{equation}
where $P_k(a_0), P_k(a_1) I_k$, with $k=1,2, ...$ are the parameters relevant to the $k-$th preparation.
%
%
\section{Risk and ambiguity in quantum formalism}
The experiments about risk and ambiguity that we  consider can always be described in
the following general way: we have an urn with red and black balls, in total number of
$N$. The agents bet on a particular color, for example black: if a black ball is drawn
from the urn, the agent wins $X$ dollars (win condition), and nothing otherwise (lose
condition). We distinguish two different bets:
\begin{itemize}
\item a clear bet, where the agents know how many red and black balls are in the urn: 
in the following we consider the case of equal number of red and black balls.
\item a vague bet, where the agents don't know how many red and black balls are in the urn: they may be in any proportion.
\end{itemize}
Our general description of risk and ambiguity effects involves the following  question: 
"the drawn ball is black?". The events relevant to this question are $b_0$ for the answer No, 
and $b_1$ for answer Yes.

In experimental tests, agents have to express their quantitative preferences about bets 
by prizing them, thus determining the cash equivalents to the bets. There are two different 
techniques to evidence such prizes: 1) the willingness to pay (WTP), that is the prize that 
an agent decides to pay  for a ticket of this lottery; 2) the willingness to accept (WTA), 
that is the prize that an agent assigns  when he sells a ticket of this lottery. 
In general, the WTP (willingness to pay) is lower than the WTA (willingness to accept) 
for the same gamble.

Of course, agents determine the cash equivalents by considering what they gain in the
win/lose condition respect to what they pay (or accept). Our hypothesis is that the
possibilities pay/not pay or accept/not accept influence the determination of the cash
equivalents (and thus the estimated probabilities to win the lottery) in the following way: we define in general the observable \textit{action}
$A$, which corresponds in the WTP to the action of pay for the ticket ($a_1$) or not
pay ($a_0$), while in the WTA to the action of accept money ($a_1$) or not accept
($a_0$).

According  to the classic Bayes's formula, the probability that a black ball is drawn
or not $P(b_i)$ (with $i=0,1$) in the clear bet can be determined by knowing the
probabilities $P(a_j)$ (with $j=0,1$)  and the conditional
probabilities $P(b_i|a_j)$: in particular, the conditional probabilities are all
$1/2$, which means that the action pay/not pay or accept/not accept can not influence
the result of drawing a black ball (they are independent events). Thus we have
$P(b_1)=P(a_0)P(b_1|a_0)+P(a_1)P(b_1|a_1)=0.5$.

According to quantum formalism, the opinion  state about $a_j$ is described, both in
the clear and in the vague bets, by the vector (in the basis $a_j$):
\begin{equation}\label{opinion_state_1}
\left [\begin{tabular}{c}
$\sqrt{P(a_0)} e^{i\phi_0}$\\
$\sqrt{P(a_1)} e^{i\phi_1}$
\end{tabular}\right]
\end{equation}
This means, since the agents do not know a-priori which  action they will choose, that
the opinion state is a linear superposition of the two possible choices. In general,
it seems reasonable to attribute  in mean the same probability to $P(a_0)=P(a_1)=1/2$: the agents evaluate the color that will be drawn by considering both the situations $a_0$ and $a_1$.
However, Bayes' formula is replaced by the interference formula (\ref{interference}),
which states
\begin{eqnarray}\label{interference_red}
& P(b_1)=P(a_0)P(b_1|a_0)+P(a_1)P(b_1|a_1)+\\
& 2\sqrt{P(a_0)P(a_1)P(b_1|a_0)P(b_1|a_1)}cos(\phi_0-\phi_1)
\end{eqnarray}
The interference equation evidences that the two beliefs about $a_j$ may interfere,
and the resulting estimated probability $P(b_1)$ may be higher or lower than $1/2$.

In the vague bet, the probability to draw a black ball $P(b_1)$ can not derived with certainty. 
However, we suppose that agents make a judgement of the probability to draw a black ball, 
by considering the two possible choices which are availble (pay/not pay in the WTP, 
accept/not accept in the WTA). Thus the conditional probabilities $P(b_i|a_j)$ are 
determined by simple arguments of statistical independence, obtaining the same equation 
of the clear bet (\ref{interference_red}).

Finally, we make the hypothesis that the presence of uncertainty situations leads to a negative 
phase factor $cos(\phi_0-\phi_1)$, thus entailing that $P(b_1)<1/2$. This fact is able to explain 
the effects of risk aversion and ambiguity aversion, as we will show in the following. 
In fact, the probability associated to the  unknown event $P(b_1)$ is lowered by the 
interference effect, even if all the agents would agree with the fact that the probability to draw a black ball is 0.5.

In figure 1 we show the interference term for three different values of $P(b_1|a_1)$
(0.25, 0.5 and 0.75), corresponding to anticorrelation/no-correlation/correlation
between $P(b_1)$ and $P(a_1)$ respectively.
\begin{figure}[h]\label{fig1}
\centering
\includegraphics[width=11.5cm]{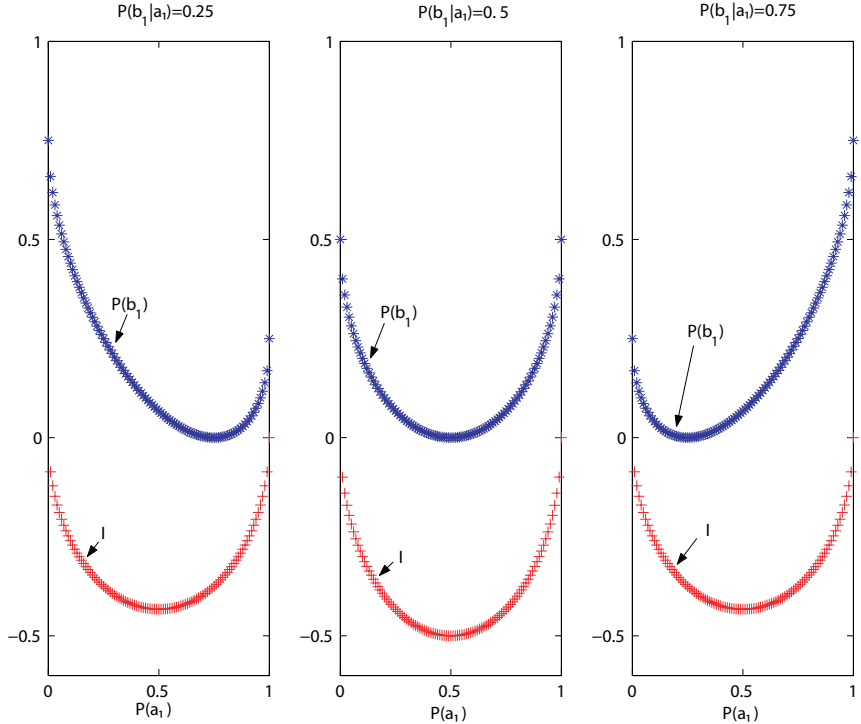}
\caption{Maximal interference term for three different values of $P(b_1|a_1)$}
\end{figure}
%
%
%
\subsection{Prizing bets}
The classic expected utility of the clear bet is defined by formula (\ref{EU}),  and thus
$$
u_c=u(b_0)P(b_0)+u(b_1)P(b_1)=X P(b_1)
$$
We consider the experimental case of \cite{Tv_Fox1}, where 100 dollars are given to the agents 
if the guessed color (black for example) is drawn, and there are 100 balls in the urn: 
$u(b_1)=100$ and $u(b_0)=0$. The classic expected payoff is calculated in the classic formalism 
(where $P(b_0)=P(b_1)=0.5$) by the following   $u(b_0)P(b_0)+u(b_1)P(b_1)=50$.

We want to  show that the quantum formalism allows us to use the same formula of the
expected utility (\ref{EU}), where the judged probabilities $P(b_i)$ are obtained by
the quantum interference formula (\ref{interference_red}), in order to compute the WTA and WTP cash
equivalents. In particular, in the simple example of \cite{Tv_Fox1}, the expected
payoff is
\begin{equation}\label{U_quant}
u=u(b_0)P(b_0)+u(b_1)P(b_1)=50 + 100 I\,.
\end{equation}
We  call the additional term the \textit{utility of uncertainty}, which is able to
modify the expected value to the extreme values of 0 or 100: the utility of
uncertainty can be interpreted as the corrective amount of money that the agents add
or subtract to the expected value in order to consider the uncertainty. The utility of
uncertainty in our example cannot be higher than $X/2$ (positive interference) or lower than $-X/2$
(negative interference).  From our hypothesis, the negative value of the phase factor
entails a negative utility of uncertainty. The numerical difference between WTA and
WTP can be explained in the quantum formalism with the presence of two different
interference terms, and thus of different utilities of uncertainty: the interference
term for WTP is higher in modulus than the interference term for WTA, since the
uncertainty to pay (and eventually to gain nothing) is perceived as more worrying.

In general, if we call $u$ the estimated cash equivalent (WTA or WTP) of a clear or vague bet, $X$ the gain in the case of drawn black ball and $u_c$ the classic expected utility, the
interference term can be obtained as
\begin{equation}\label{I}
I=\frac{u-u_c}{X}
\end{equation}
If the opinion state of all the decision makers is identical, the cash equivalent is
the same, leading to the same preferences for all the agents. However, a more
realistic situation requires the use of a mixed state, as in formula
(\ref{interference_mixed}). The mean prize for a bet is determined by the percentage
of agents which set a given prize. In formula, the expected utility for the clear and
vague bets previously defined is
$$
u=\sum_k P_k  u_k=\sum_k P_k X P_k(b_1)
$$
where $P_k$ is the probability to find an agent in the preparation $k$, while
$P_k(b_1)$ is the estimated probability of event $b_1$ in the preparation $k$. The
mixed state of two different opinion states about the same bet with different
utilities of uncertainty evidences a final utility of uncertainty which is comprised
between these two utilities of uncertainty.
\subsection{Risk effects}
Given the clear bet described before, if the mean cash equivalent is less/equal/higher
than the expected  outcome of the same lottery, we have a situation of \textit{risk
avoidance/neutrality/seeking} \cite{Curley_Yates}. In the quantum formalism, this
implies that for the clear bet the utility of uncertainty is negative/null/positive
respectively.

In the study I of \cite{Tv_Fox1} the mean WTP for the clear bet,  evaluated alone, is
17.94, which is lower than 50 (the classic expected outcome). Thus we have a situation of risk
avoidance. By using equation (\ref{U_quant}), we can describe this fact by noting that
the cash equivalent is $50-100I=17.94$, obtaining $I=-0.3$. We can compare this
experimental result with figure 1. The estimated conditional probability
$P(b_1|a_1)=0.5$ allows us to consider the middle part of the central panel in figure
1. The maximum interference effect in this case is  about $-0.5$, which is higher in
modulus than -0.3. The presence of a mixed state is sufficient to explain such a
difference. Some information about the composition of the mixed state can be obtained
by considering the standard error, which is 2.5: this is consistent with a high
percentage of agents with highly negative interference terms (risk avoidance), and a
low percentage of risk seeking/neutral agents.

In the study II of \cite{Tv_Fox1} (table 2) a similar test is performed by considering
urns with only two ping ping balls and by asking for the WTA. Given the winning payoff
20\$, the classic expected utility is 10\$. Of couse the WTA measure leads to higher
values than WTP, which entails less strongly negative interference terms. We consider
in this subsection only the clear bet evaluated alone (noncomparative case), which has
a WTA of  7.58 ($I=-0.121$),

Finally, another feature influencing  the interference term is the payoff. In
\cite{Halevy}, the two-color Ellsberg experiment is performed, and the same clear bet
is tested with a low payoff (2 dollars) and a high payoff (20 dollars). The mean WTA in the low case
is 1.061, evidencing a weak risk seeking effect, while in the high case is 8.37,
leading to a men interference term of $-0.163$.
%
%
\subsection{Ambiguity}\label{sect:ambiguity}
The difference between risk and ambiguity  is that risk can be expressed by precise probabilities, while ambiguity  (also called unmeasurable uncertainty) cannot.
Interest in the concept of ambiguity was revived by Ellsberg \cite{ellsberg}, who
showed that people generally prefer to bet on known rather than unknown probabilities.
Ambiguity aversion seems to represent a reluctance to act on inferior knowledge, and
this inferiority is brought to mind only through a comparison with superior knowledge
about other domains or of other people. The Ellsberg's simplest example, known as the
\textit{two-color problem}, clearly shows the effect of ambiguity aversion. It
involves the same experimental apparatus described before: urn 1 contains 50 red and
50 black balls, whereas urn 2 contains 100 red and black balls in an unknown
proportion. We suppose that a ball is drawn at random from an urn and one receives
\$100 or nothing depending on the outcome. Most people seem indifferent between
betting on red or on black for either urn, yet they prefer to bet on the 50-50 urn
rather than on the urn with the unknown composition. This pattern of preferences is
inconsistent with expected utility theory because it implies that the subjective
probabilities of black and of red are greater in the 50-50 urn than in the unknown
urn, and therefore cannot sum to one for both urns.
%
%
%
\subsection{The order effects and the quantum gates}
Clear and vague bets, when evaluated alone, do not exhibit significantively different WTA or WTP. This can be shown in \cite{Tv_Fox1}, study 1 and 2 (non-comparative condition), or in the table below, where the interference term has been calculated.  In particular, we note that the interference term is higher in modulus in the WTP, since this kind of choice is perceived as more risky. Moreover, we note that the interference term for the clear bet is higher in modulus than for the vague of about 0.02 (thus they are almost equal). 
\\\\
\begin{tabular}{|c||c|c|}\hline
 Non-comparative    & Study 1 of \cite{Tv_Fox1} (100 balls, 100\$) & Study 2 of \cite{Tv_Fox1} (2 balls, 20\$) \\\hline\hline
   clear     & WTP=17.94, $I=-0.32$ & WTA=7.58, $I=-0.12$  \\\hline
   vague     & WTP=18.42, $I=-0.31$ & WTA=8.04, $I=-0.10$  \\\hline
\end{tabular}
\\\\
However, when the two bets are evaluated toghether by the same agents (the comparative condition), the ambiguity aversion effect appears. In the table below we evidence the results of \cite{Tv_Fox1}, study 1 and 2 (comparative case), with the interference terms. However, in this experiment the order in which the two bets were presented was counterbalanced.
\\\\
\begin{tabular}{|c||c|c|}\hline
  Comparative   & Study 1 of \cite{Tv_Fox1} (100 balls, 100\$) & Study 2 of \cite{Tv_Fox1} (2 balls, 20\$) \\\hline\hline
   clear     & WTP=24.34, $I=-0.26$ & WTA=9.74, $I=-0.01$  \\\hline
   vague     & WTP=14.85, $I=-0.35$ & WTA=8.53, $I=-0.07$  \\\hline
\end{tabular}
\\\\
Thus we focus our attention on the order effects in the evaluation of
cash equivalents: in study 3 of \cite{{Fox_Weber}} (with the same experimental
description of the study 2 of \cite{Tv_Fox1}), the WTA's in the order "clear and then vague" are 8.92 and 7.50, while in the order "vague and then clear" they are 9.56 and 10.56. This evidences that the presence of a vague bet after a clear bet leads to a high perceived ambiguity. On the contrary, if we evaluate first a vague and then a clear bet, no ambiguity is perceived at first, and then ambiguity diminishes, leading to higher WTA.

We try to explain this effect consistently with the quantum formalism by making the hypothesis that the opinion state of agents, while evaluating the two bets in the defined order, can be described by a single qubit. In the evaluation of the first bet, we have a similar opinion state both in the clear and vague bets, as noted before (non-comparative situation). However, when the second bet is introduced, we make the hypothesis that a suitable unitary operator is applied on the previous opinion state: this can be interpreted as the effect of the information given with the second bet.

The unitary operator we will use is a phase rotation gate $R$ (or phase shift single-qubit gate)
\begin{equation}\label{rotation}
R(\xi)=\left[
    \begin{tabular}{cc}
    1 & 0 \\
    0 & $e^{i\xi}$
    \end{tabular}\right]\,\,,
\end{equation}
whose action is on the single qubit describing the opinion state. The opinion state
after the phase shift is
\begin{equation}\label{phase_shift}
\left[
    \begin{tabular}{cc}
    1 & 0 \\
    0 & $e^{i\xi}$
    \end{tabular}\right]
\left[
    \begin{tabular}{c}
    $\frac{1}{\sqrt{2}}e^{i\phi_0}$\\
    $\frac{1}{\sqrt{2}}e^{i\phi_1}$
    \end{tabular}\right]
=
\left[
    \begin{tabular}{c}
    $\frac{1}{\sqrt{2}}e^{i\phi_0}$\\
    $\frac{1}{\sqrt{2}}e^{i(\phi_1+\xi)}$
    \end{tabular}\right]
\,.
\end{equation}
Thus the new interference term for the second bet contains  the phase factor
$cos(\phi_1-\phi_0+\xi)$, leading to a more/less negative utility of uncertainty. We note that if $cos(\phi_1-\phi_0)$ is near to zero and $\xi$ is very close to zero, we can use the linear approximation $cos(\phi_1-\phi_0+\xi)=\pi/2 -(\phi_1-\phi_0+\xi)$. In the table below we rewrite the results of study 3 of \cite{{Fox_Weber}}, with the values of the interference term and of the phase shift $\xi\simeq 2(I_2-I_1)$:
 \\\\
\begin{tabular}{|c||c|c|}\hline
     & clear $\rightarrow$ vague & vague $\rightarrow$ clear   \\\hline\hline
   1     & WTA=8.92, $I=-0.05$ & WTA=9.56, $I=-0.02$  \\\hline
   2     & WTA=7.50, $I=-0.13$ & WTA=10.56, $I=0.03$  \\\hline\hline
         & $\xi\simeq 0.16$ & $\xi\simeq -0.14$  \\\hline
\end{tabular}
%
%
\subsection{The comparative ignorance hypothesis}
The comparative ignorance hypothesis, first developed in \cite{Tv_Fox1}, states that ambiguity aversion will be present when subjects evaluate clear and vague prospects jointly, but it will greatly diminish or disappear when they evaluate each prospect in isolation.  This is consistent with the results of the study I of \cite{Tv_Fox1}, where the mean WTP relevant to the vague bet, evaluated alone, is very similar to the WTP relevant to the clear bet, evaluated alone. This situation is named the \textit{noncomparative} case. The \textit{comparative} case is the situation where the clear and the vague bets are evaluated by the same agents in the same test (the order in which the two bets were presented was counterbalanced). The surprising result is that now the WTP for the vague bet is much lower than for the  clear bet.
A similar result is obtained in the study 2 of \cite{Tv_Fox1} (table 2) where  two urns contain only two ping ping balls, and the cash equivalent is the WTA. 

We want to show that in the comparative case the WTA are exactly the mean of the corresponding WTA in the two possible evaluation orders. This is consistent with the interpretation that about half of the agents evaluate with the order clear/vague, and the other half with the opposite order vague/clear. The WTA of the vague bet in \cite{Tv_Fox1} is 8.53, which is exactly equal to the mean of the values $(7.50+9.56)/2$ from \cite{{Fox_Weber}}. In analogous way, we have for the clear bet $9.74=(8.92+10.56)/2$
%
%
%

In \cite{Halevy}, the two-color Ellsberg experiment is performed in a different way. First of all the urn contains 10 balls, red or black, and the cash equivalent is WTA. Two versions of the experiment are presented, one with the winning payoff equal to 2\$, and the other with 20\$. The clear bet consists in 5 red and 5 black balls, while the vague bets are designed in different ways: a) The number of red and black balls is unknown, it could be any number between 0 red balls (and 10 black balls) to 10 red balls (and 0 black balls). b) The number of red and black balls is determined as follows: one ticket is drawn from a bag containing 11 tickets with the numbers 0 to 10 written on them. The number written on the drawn ticket will determine the number of red balls in the third urn. For example, if the ticket drawn is 3, then there will be 3 red balls and 7 black balls. c) The color composition of balls in this urn is determined in a similar to before. The difference is that instead of 11 tickets in the bag, there are 2, with the numbers 0 and 10 written on them. Therefore, the urn may contain either 0 red balls (and 10 black balls) or 10 red balls (and 0 black balls).
In the table below we show the measured mean WTA, and the relevant interference terms:
\\\\
\begin{tabular}{|c||c|c|c|c|}\hline
     & Clear & Vague a) & Vague b) & Vague c) \\\hline\hline
   2\$     & WTA=1.061, I=0.03 & 0.878/ I=-0.06 & 0.929/ I=-0.0355& 0.948/  I=-0.026 \\\hline
   20\$    & WTA=8.37, I=-0.0815 & 6.66/ I=-0.167& 7.25/ I=-0.137 & 7.74/ I=-0.113  \\\hline
\end{tabular}
\\\\
The interference terms are more negative in the 20\$ case, evidencing a stronger perception of risk and ambiguity. Moreover, the vague case c) shows interference terms lower in modulus than the other vague cases: this can be explained by the fact that the presence of only two possibilities diminishes the ambiguity of the bet and thus the modulus of the uncertainty utility. Vague bets a) and b) are very similar, but the case a) has interference terms higher in modulus. In fact in b) the mechanism  to determine the number of red/black balls seems to help to reduce the perceived ambiguity. 
\section{Conclusion}
We have developed within the quantum formalism a simple model which is able to describe the effects of risk and ambiguity on decisions: in particular, we have used the formula of quantum interference, which leads to an additional term in the computation of the expected utility. The interpretation of such additional term, called the \textit{uncertainty utility}, is the utility equivalent to the uncertainty perceived by the agents.
Such term is in general negative (but in some cases it can be positive), and its absolute value depends on the following factors: 1) it increases with the vagueness, when compared with a clear bet, 2) it increases with the expected payoff, since the choice is perceived as more worring, 3) it is higher for WTA than for WTP.
Thus the uncertainty utility, or the interference term, manifests a psychological nature. 

Open questions which remain to be studied are:
a) the link between prospect theory and quantum formalism, 
b) the connection with works about order effects in belief updating, 
c) the relation of the present formalism with other quantum-like approaches such as \cite{Mura1}, and
d) an axiomatic approach consistent with quantum interference effects.
 \footnotesize
%
%


\section*{References}

\begin{thebibliography}{99}
%
%
\bibitem{Rfranco_rat_ign_3} R. Franco, The inverse fallacy and quantum formalism
http://xxx.lanl.gov/abs/0708.2972v1
%
\bibitem{Rfranco_rconj} R. Franco, The conjunction fallacy and interference effects, http://xxx.lanl.gov/pdf/0708.3948
%
%
\bibitem{Mura1} Pierfrancesco La Mura, Decision Theory in the Presence of Uncertainty and Risk, Quelle: HHL-Arbeitspapier Nr. 68, 11 S (2005)
%
\bibitem{Neumann} John von Neumann and Oskar Morgenstern (1944), Theory of Games and
Economic Behavior. Princeton University Press, Princeton NJ.

\bibitem{ellsberg} Ellsberg, Daniel, "Risk, Ambiguity and the Savage Axioms," Quarterly Journal of Economics, LXXV (1961), 643-69.
\bibitem{Keynes} Keynes, John Maynard, A Treatise on Probability (London: Macmillan, 1921).
\bibitem{Tv_Fox1} Fox, C.R., and Tversky, A. (1995).  Ambiguity aversion and comparative ignorance.  Quarterly Journal of Economics, 110, 585-603.
\bibitem{Fox_Weber} Fox, C.R., and Weber, M.  (2002).  Ambiguity aversion, comparative ignorance, and decision context. Organizational Behavior and Human Decision Processes, 88, 476-498.
%
\bibitem{Curley_Yates} S. P. CURLEY, J. F. YATES, and A. A. ABRAMS, Psychological Sources of Ambiguity Avoidance, Organizational behavior and  human decision processes 38, 230-256 (1986)
%
\bibitem{Sarin_Weber} Sarin, R. K., \& Weber, M. (1993). Effects of Ambiguity in Market
Experiments. Management science, 39(5), 602
%
\bibitem{Ho_Keller}
Ho, J. L. Y., Keller, L. R., \& Keltyka, P. (2002). Effects of
Outcome and Probabilistic Ambiguity on Managerial Choices. Journal
of Risk and Uncertainty, 24(1), 47-74.
%
%
\bibitem{Halevy} Y. Halevy,  Ellsberg Revisited: An Experimental Study, Econometrica (2007), 75 (2), 503–536. Supplemental Material
%
%
%
%
\end{thebibliography}
\end{document}